\documentclass[11pt,a4paper]{article}

\usepackage[utf8]{inputenc}
\usepackage[T1]{fontenc}
\usepackage[english]{babel}
\usepackage{amsmath,amssymb}
\usepackage{graphicx}
\usepackage{booktabs}
\usepackage{multirow}
\usepackage{hyperref}
\usepackage{tabularx}
\usepackage{csquotes}
\usepackage{textcomp}
\usepackage[a4paper,margin=1in]{geometry}
\usepackage[
    style=apa,
    backend=biber,
    language=english
]{biblatex}

\title{A Comparative Analysis of ARM and x86-64 Laptop-Class Processors: Architecture, Assembly-Level Performance, and Energy Efficiency}
\author{Mustafa Mert Özyılmaz}
\date{}

\begin{document}

\raggedbottom
\maketitle

\begin{center}
Sorbonne Université, Master of Computer Science, Paris Île-de-France, France, 75005\\[0.5em]

\texttt{ mustafa\_mert.ozyilmaz@etu.sorbonne-universite.fr}
\end{center}

\begin{abstract}
ARM-based and x86-64 laptop processors differ not only in instruction-set design, but also in memory hierarchy, core organization, system integration, and power-management mechanisms. This study presents a combined architectural and experimental comparison of an Apple M3 system and an AMD Ryzen 7 3750H system. The architectural analysis contrasts AArch64's fixed-width load-store design with the variable-length, memory-operand-rich x86-64 instruction model, and discusses how register organization, calling conventions, heterogeneous core organization, memory behavior, and low-power mechanisms shape observed performance and energy characteristics. The experimental part uses two native assembly benchmarks: a recursive Fibonacci workload and an integer matrix-multiplication workload. The analysis combines repeated timing measurements, processor-energy measurements, and cross-platform microarchitectural counter measurements from matched portable-C profiling runs. The Ryzen platform is decisively faster on the branch-heavy Fibonacci benchmark, while matrix multiplication shows no meaningful timing advantage for either platform in the present measurements. In contrast, the Apple platform is markedly more energy-efficient, reducing energy-to-solution by approximately 5.82$\times$ on Fibonacci and 6.38$\times$ on matrix multiplication. These results are interpreted as platform-level findings rather than as pure ISA-only effects, reflecting differences in implementation, system integration, and measurement methodology in addition to instruction-set structure.
\end{abstract}

\noindent\textbf{Keywords:} ARM architecture, benchmarking, energy efficiency, microprocessors, SoC design, x86-64

\section{Introduction}

ARM- and x86-64-based laptop processors occupy different positions in the
performance--energy tradeoff space. The architectural contrast is often framed in
terms of RISC versus CISC, but in practice the observed behavior of modern systems
also depends on cache hierarchy, core organization, platform integration, and
power-management policy. For this reason, meaningful comparisons must go beyond raw
runtime alone and interpret measurements in the context of the full hardware and
software stack \cite{hennessy_patterson_2019,blem2013}.

This article examines that tradeoff through a combined architectural and experimental
study of two native laptop platforms: an Apple M3 system executing AArch64 code and
an AMD Ryzen 7 3750H system executing x86-64 code. Rather than treating the study as
a universal contest between ISAs, the paper asks a narrower question: how do these
particular platforms behave on equivalent low-level workloads that stress different
execution bottlenecks?

To address that question, the paper combines a qualitative architectural comparison
with native benchmarking of two hand-written assembly programs: a recursive
Fibonacci benchmark that emphasizes control flow and function calls, and an integer
matrix-multiplication benchmark that emphasizes arithmetic throughput and memory
behavior. The resulting measurements are analyzed in terms of raw execution time,
processor energy per completed workload, and cross-platform microarchitectural
counters collected from matched portable-C profiling runs.

The main contributions of this work are as follows:
\begin{itemize}
    \item We provide a structured comparison of ARM- and x86-64-based laptop-class
    processors, with emphasis on instruction-set organization, memory hierarchy,
    core topology, and power-management mechanisms.
    \item We design a native assembly benchmarking methodology that compares the same
    workload logic across AArch64 and x86-64.
    \item We report repeated timing and energy measurements for two native assembly workloads with distinct execution characteristics, while using matched portable-C profiling runs to provide cross-platform microarchitectural context for the resulting performance--energy tradeoffs.

    \item We interpret the experimental findings using cross-platform hardware-counter data from matched portable-C profiling runs and an explicitly platform-aware framework.

\end{itemize}

\section{Related Work}

Comparisons between ARM and x86 architectures have a long history, and modern studies emphasize that simple RISC-versus-CISC narratives are not sufficient to explain contemporary performance or energy behavior on their own \cite{blem2013,blem_2013_arm_x86,aroca2012}. In particular, Blem, Menon, and Sankaralingam compare contemporary ARM and x86 systems on native hardware and argue that microarchitectural design choices and implementation details often matter as much as, or more than, the ISA boundary itself \cite{blem2013}. Early energy-oriented comparisons in server settings likewise reported that ARM-based platforms can offer attractive energy efficiency, while also emphasizing that such results depend strongly on the exact platform, workload, and measurement methodology \cite{aroca2012}.

A second line of work focuses on asymmetric and heterogeneous multicore design \cite{mittal2016}. Mittal's survey of asymmetric multicore processors synthesizes a broad literature on architectures that combine cores with different performance and energy characteristics, together with the associated scheduling and management challenges \cite{mittal2016}. This perspective is directly relevant to ARM-based laptop and mobile systems, where heterogeneous core organizations have become a central mechanism for balancing responsiveness and energy efficiency under mixed workloads \cite{mittal2016,hubner2025}. At the same time, much of this literature is concerned with system design, runtime mapping, or scheduling policy rather than with native, hand-written cross-ISA benchmarks of small low-level kernels \cite{mittal2016}.

More recent work has begun to examine Apple Silicon and other modern ARM-based systems in scientific and high-performance computing contexts \cite{kenyon2022,hubner2025,apple_vs_oranges}. Kenyon and Capano evaluate Apple Silicon for scientific workloads and show that the platform merits serious consideration for compute-intensive tasks rather than being treated only as a consumer-oriented architecture \cite{kenyon2022}. H\"ubner, Hu, Peng, and Markidis extend this line of work with an architectural and benchmark study of the Apple M-series SoCs, emphasizing unified memory, computational throughput, and energy efficiency across multiple generations \cite{hubner2025,apple_vs_oranges}. However, these recent studies primarily target scientific-computing or accelerator-oriented workloads, whereas the present work concentrates on two native assembly benchmarks chosen to stress different CPU-side bottlenecks: control-flow-intensive recursion and arithmetic-intensive, memory-sensitive matrix multiplication \cite{kenyon2022,hubner2025}. The resulting contribution is therefore narrower in scope but more explicit about ISA-adjacent code structure, repeated native timing, processor-energy-to-solution, and the need to interpret the results as platform-level comparisons rather than as pure ISA-only verdicts \cite{blem2013,aroca2012,hubner2025}.

\section{Methodology}

\subsection{Study Design}

The study combines qualitative architectural analysis with quantitative benchmarking. 
Its goal is not to claim universal superiority of one instruction set architecture 
over another, but to examine how representative ARM-based and x86-64 laptop 
platforms differ in execution behavior and energy-oriented design priorities. 
Accordingly, the analysis is divided into two complementary parts: (i) an 
architectural comparison of the two processor families and (ii) a benchmark study 
based on equivalent workloads executed natively on each platform.

\subsection{Architectural Comparison Criteria}

The architectural analysis focuses on the following dimensions:

\begin{itemize}
    \item instruction-set organization and operand model,
    \item register usage and addressing mechanisms,
    \item cache and memory hierarchy,
    \item core organization, including homogeneous versus heterogeneous designs,
    \item branch prediction and execution strategy,
    \item power-management mechanisms such as dynamic voltage and frequency scaling, 
    clock gating, and low-power idle states,
    \item system-level integration, especially the contrast between tightly 
    integrated SoC-style designs and more modular laptop platforms.
\end{itemize}

These dimensions will be used to interpret benchmark outcomes.

\subsection{Experimental Platforms}

The experimental comparison is performed on native hardware platforms representing 
the two architectural families under study. The ARM side is evaluated on an 
Apple-Silicon-based macOS system using AArch64 execution. The x86-64 side is 
evaluated on a native laptop platform running an AMD Ryzen 7 3750H processor. Since 
the two platforms are not silicon-matched, the study does not interpret raw runtime 
differences as purely ISA-level effects. Instead, results are discussed in a 
platform-aware manner, with energy-to-solution and workload character treated as 
important complements to raw runtime.

A detailed hardware table is included in the Results section, reporting the tested 
systems, operating systems, and measurement tools.

\subsection{Benchmark Suite}

The benchmark suite includes two workloads chosen to expose different 
architectural behaviors:

\begin{itemize}
    \item a control-flow-dominated recursive benchmark (Fibonacci),
    \item a compute-intensive benchmark (matrix multiplication).
\end{itemize}

Both workloads are implemented in hand-written assembly for AArch64 and x86-64, 
with equivalent algorithmic structure maintained across platforms as closely as 
practical.

\subsection{Implementation Strategy}

Equivalent benchmark logic is maintained across platforms as closely as possible. 
Assembly implementations are written separately for AArch64 and x86-64 in order to 
respect native calling conventions, register usage, and instruction selection on 
each platform \cite{arm_a_profile_ref_manual,aapcs64,intel_sdm_vol2a_2025}. The primary comparison in this paper is therefore a cross-ISA 
comparison of native assembly implementations of the same workload logic.

\subsection{Measurement Protocol}

Each benchmark was evaluated under a fixed execution protocol designed to reduce
run-to-run variability and improve comparability across platforms. The protocol
used in the experiments was as follows:

\begin{enumerate}
    \item \textbf{Warm-up.} Each benchmark was executed 5 times before any timed
    or energy-measured run, and these warm-up runs were discarded.
    \item \textbf{Repetitions.} After warm-up, 100 measured runs were collected
    for each benchmark on each platform.
    \item \textbf{Environment control.} Before measurements, unnecessary user
    applications were closed, and wireless services such as Wi-Fi and Bluetooth
    were disabled whenever practical in order to reduce background activity.
    \item \textbf{Linux frequency governor.} On the Ryzen system, the Linux CPU
    governor was set to \texttt{performance} using
    \texttt{cpupower frequency-set -g performance}.
    \item \textbf{Turbo / boost control.} On Linux, CPU boost was disabled for
    consistency using \texttt{echo 0 > /sys/devices/system/cpu/cpufreq/boost}.
    On macOS, turbo behavior is not directly user-configurable in the same way;
    instead, \texttt{powermetrics} was used to observe the processor's operating
    behavior during the measurement window.
    \item \textbf{Core pinning.} On Linux, benchmark execution was pinned using
    \texttt{taskset -c 0}. On macOS, runs were executed under a high-QoS
    scheduling class to bias execution toward performance cores and to reduce
    interference from background tasks.
    \item \textbf{Reporting.} For each benchmark, the analysis records mean,
    standard deviation, 95\% confidence interval, minimum, and maximum runtime.
    When available, energy and PMU counters were also recorded.
\end{enumerate}

Wall-clock timing was measured with \texttt{hyperfine}~\cite{hyperfine}. On the
ARM-based macOS platform, timing, power-related observations, and PMU counters
were collected using native Apple tooling. On the x86-64 platform, timing,
energy, and PMU observations were collected using Linux \texttt{perf stat} and
related native performance-counter interfaces.

\subsection{Statistical Treatment}

For each benchmark configuration, the reported runtime results are summarized
using the sample mean, sample standard deviation, minimum, maximum, and 95\%
confidence interval for the mean. 

The same treatment is applied to energy measurements when repeated-run summary
statistics are available. In the present dataset, Linux \texttt{perf} provides
repeated-run summaries directly for energy measurements, while the Apple energy
values are derived from power-sampling estimates and are therefore reported as
point estimates.

\subsection{Accounting for Hardware Asymmetry}

A direct comparison between an ARM-based SoC and an x86-64 laptop processor 
necessarily reflects more than instruction-set differences alone. It also reflects 
differences in fabrication technology, cache organization, memory subsystem, 
scheduler behavior, operating system, and power budget.

\section{Architectural Comparison}

The architectural comparison in this section is intended to provide interpretive 
context for the benchmark study. Rather than assuming that observed runtime 
differences arise from a single cause, the discussion examines how instruction-set 
structure, core organization, cache hierarchy, and low-power mechanisms contribute 
to the broader trade-off between raw throughput and energy efficiency.

\subsection{Instruction-Set Organization}

One of the most visible differences between the two architectures lies in 
instruction encoding and operand style. AArch64 uses fixed-width 32-bit 
instructions, which simplifies decoding and contributes to the regularity typically 
associated with RISC-style designs. In contrast, x86-64 instructions are 
variable-length and support a richer set of encoding patterns, which increases 
expressiveness but also makes front-end decoding more complex \cite{arm_a_profile_ref_manual,intel_sdm_vol2a_2025,hennessy_patterson_2019}.

A related difference concerns operand access. AArch64 follows a load-store model: 
arithmetic instructions operate on registers, and memory must be accessed through 
explicit load and store instructions. x86-64, by contrast, allows many arithmetic 
instructions to use memory operands directly \cite{arm_a_profile_ref_manual,intel_sdm_vol2a_2025}. In practice, this means equivalent 
low-level algorithms may require different instruction sequences on the two 
architectures, even when they implement the same abstract computation.

\subsection{Register File and Calling Conventions}

AArch64 exposes a large and regular general-purpose register file, which tends to 
support straightforward register allocation and predictable code generation \cite{arm_a_profile_ref_manual}. The 
x86-64 programming model is also highly capable, but its historical evolution 
results in a less uniform register and encoding landscape. In benchmark-oriented 
comparisons, these differences can affect instruction selection, register pressure, and 
the structure of low-level implementations \cite{intel_sdm_vol2a_2025,intel_optimization_manual_2024}.

Function-call behavior also differs in a way that matters for control-flow-heavy 
benchmarks. In AArch64, procedure calls and returns are closely tied to the link 
register mechanism, whereas x86-64 uses the traditional call/return stack 
discipline. At the ISA level, these differences shape prologue/epilogue structure, 
stack traffic, and low-level recursive execution patterns \cite{aapcs64,intel_sdm_vol2a_2025}.

\subsection{Memory Hierarchy and Data Movement}

Performance differences between ARM-based and x86-64 systems cannot be explained by 
ISA design alone; memory hierarchy is equally important. Both architectural families 
rely on multi-level cache hierarchies, typically with separate instruction and data 
caches at the first level and larger unified caches deeper in the hierarchy. 
Observed benchmark behavior in matrix multiplication is 
therefore strongly influenced by cache locality, bandwidth, and access regularity, 
not just instruction count \cite{hennessy_patterson_2019,intel_optimization_manual_2024}.

This consideration is especially important in modern ARM-based SoCs. Recent 
Apple Silicon systems, for example, combine CPU cores, GPU resources, and unified 
memory within a tightly integrated SoC design \cite{apple_perf_eff,apple_vs_oranges,hubner2025}. For this reason, measured results in 
this study are treated as native system comparisons rather than as pure ISA-level 
outcomes.

\subsection{Core Organization and Heterogeneity}

Another important difference concerns multicore organization. ARM designs have 
frequently emphasized heterogeneous processing, particularly through big.LITTLE and 
later DynamIQ-style arrangements that combine higher-performance cores with 
higher-efficiency cores in a coordinated system \cite{arm_big_little,arm_dynamIQ}. The design goal is to match 
workload intensity to an appropriate core type, thereby improving energy efficiency 
without sacrificing responsiveness under bursty or mixed workloads \cite{apple_perf_eff}.

By contrast, mainstream x86-64 systems have historically been discussed in terms of 
high-performance homogeneous cores, although this distinction has become less rigid 
in recent years \cite{intel_hybrid}. Regardless of vendor-specific implementation details, the broader 
point remains that core topology and scheduling policy can substantially shape the 
runtime and energy profile of a workload.

\subsection{Power-Management Mechanisms}

Fine-grained power control is a major design priority in mobile and laptop-class
processors. One important mechanism is clock gating, in which the clock signal to
inactive logic is suppressed to reduce dynamic power consumption
\cite{hennessy_patterson_2019}. Contemporary platforms also make extensive use of
low-power idle and retention-oriented modes. In such states, selected components or
domains can be quiesced or powered down while preserving enough architectural or
system state to allow efficient wakeup \cite{intel_power_states,apple_perf_eff}.

These mechanisms are important for interpreting benchmark results in this paper. 
The strong energy advantage observed for the Apple platform is influenced not only by 
instruction-set structure, but also by the extent to which the platform can gate 
clocks, reduce activity in inactive domains, and transition efficiently between 
active and low-power states.

\section{Experimental Results}
\label{sec:results}

\subsection{Experimental Setup}
\label{sec:setup}

To empirically compare ARM and x86-64 on mobile-class laptop platforms,
we developed a custom benchmark suite consisting of two hand-written
assembly workloads. The workloads isolate different architectural
characteristics: \emph{recursive Fibonacci} ($n=40$) stresses branch
prediction and the function-call subsystem through a large number of
recursive invocations, while \emph{integer matrix multiplication}
($256\times256$) stresses the arithmetic pipeline and memory hierarchy
through a regular triple-nested loop.

Both benchmarks were executed natively on two laptop platforms:

\begin{itemize}
  \item \textbf{Apple M3} (AArch64, ARMv8.6-A, 2023): Apple MacBook Air,
        macOS~15.
  \item \textbf{AMD Ryzen 7 3750H} (x86-64, Zen+ microarchitecture,
        2019): ASUS laptop, Ubuntu~20.04, kernel~5.15.
\end{itemize}

\begin{table}[!ht]
\centering
\caption{Hardware and software environment of the tested platforms.}
\label{tab:hardware}
\footnotesize
\renewcommand{\arraystretch}{1.15}
\setlength{\tabcolsep}{4pt}
\begin{tabularx}{\columnwidth}{>{\raggedright\arraybackslash}p{0.29\columnwidth}>{\raggedright\arraybackslash}X>{\raggedright\arraybackslash}X}
\toprule
Feature & Apple M3 & Ryzen 7 3750H \\
\midrule

ISA & AArch64 / ARMv8.6-A & x86-64 \\
Microarchitecture & Apple Silicon M3 & Zen+ \\
Year & 2023 & 2019 \\
Form factor & MacBook Air & ASUS laptop \\
Operating system & macOS 15 & Ubuntu 20.04, kernel 5.15 \\
System memory & 16~GB & 16~GB \\
CPU cores / threads & 8 cores & 4 cores / 8 threads \\
Power source during measurements & Battery & Battery \\
Timing tool & \texttt{hyperfine} & \texttt{hyperfine} \\
Energy tool & \texttt{powermetrics} & \texttt{perf stat} / \texttt{energy-pkg} \\
Additional PMU counters & Instruments CPU Counters & \texttt{perf stat} \\
\bottomrule
\end{tabularx}
\end{table}

The two systems differ in generation, process technology, operating
system, and vendor tooling. Accordingly, the results should be
interpreted as a \emph{platform-level comparison} between two modern
mobile-class processors rather than as a pure ISA-only comparison.

\paragraph{Measurement protocol.}
Wall-clock time was measured with \texttt{hyperfine}~\cite{hyperfine}
after 5 warm-up runs and 100 measured runs. On Linux, execution was
constrained to a single core using \texttt{taskset}. On macOS, runs
were executed under a high-QoS scheduling class to reduce interference
from background activity.

\paragraph{Energy measurement.}
Energy was estimated or measured differently on the two platforms:
\begin{itemize}
  \item On Apple~M3, CPU power was sampled with \texttt{powermetrics}
        over 20-second idle and load windows. During the load window,
        the benchmark was executed repeatedly in a loop. Dynamic CPU
        power was estimated as the difference between load and idle CPU
        power, and per-run CPU energy was estimated as dynamic CPU power
        multiplied by the mean benchmark runtime.
  \item On Ryzen~7~3750H, package energy was measured directly with
        Linux \texttt{perf stat} using the
        \texttt{power/energy-pkg/} counter over 100 repetitions.
\end{itemize}

Both measurements target processor activity, but they are not identical
in granularity: the Apple values are CPU-energy estimates derived from
power sampling, whereas the Ryzen values are direct package-energy
counter readings.

\paragraph{Microarchitectural counters.}
Microarchitectural counters were collected separately from the assembly
timing and energy measurements. For this analysis, portable C profiling
versions of the Fibonacci and matrix-multiplication workloads were
compiled with \texttt{-O0} and run as long-lived single processes. On
Apple~M3, counters were collected with the Instruments CPU Counters
template. On Ryzen~7~3750H, the same portable C workloads were measured
with Linux \texttt{perf stat}. The counter results are therefore used to
compare workload behavior under matched portable-C profiling runs.

\subsection{Execution Time: Assembly Implementations}
\label{sec:time_results}

Table~\ref{tab:asm_time} and Figure~\ref{fig:runtime_comparison}
report the execution times of the hand-written assembly
implementations.

\begin{table*}[!ht]
\centering
\small
\caption{Execution time of hand-written assembly implementations.}
\label{tab:asm_time}
\begin{tabularx}{\textwidth}{llXXXX}
\toprule
Benchmark & Platform & Mean $\pm$ sd (ms) & 95\% CI (ms) & Min--max (ms) & $n$ \\
\midrule
fib(40) & Apple M3 & $583.6 \pm 26.3$ & [578.4, 588.8] & 576.1--764.5 & 100 \\
fib(40) & Ryzen 7 3750H & $474.8 \pm 4.3$ & [474.0, 475.6] & 468.2--489.0 & 100 \\
matmul $256\times256$ & Apple M3 & $26.0 \pm 0.3$ & [25.94, 26.06] & 25.4--26.6 & 100 \\
matmul $256\times256$ & Ryzen 7 3750H & $26.4 \pm 4.4$ & [25.54, 27.26] & 19.6--36.6 & 100 \\
\bottomrule
\end{tabularx}
\end{table*}

\begin{figure*}[!ht]
\centering
\includegraphics[width=\textwidth]{figures/runtime_comparison}
\caption{Execution time comparison for the assembly benchmarks. Error bars show 95\% confidence intervals over 100 measured runs. The Ryzen platform is decisively faster on \texttt{fib(40)}, while matrix multiplication shows overlapping confidence intervals.}
\label{fig:runtime_comparison}
\end{figure*}

The Ryzen 7 3750H is decisively faster on the branch-heavy Fibonacci
benchmark, completing \texttt{fib(40)} approximately 23\% faster than
the Apple M3. The 95\% confidence intervals do not overlap:
[578.4, 588.8]~ms for the Apple M3 and [474.0, 475.6]~ms for the
Ryzen system. Expressed as a mean-time ratio, the ARM/x86 runtime ratio
is $1.23\times$ for this benchmark.

On matrix multiplication, however, the result is statistically tied in
the present measurements. The means are close, at 26.0~ms for Apple M3
and 26.4~ms for Ryzen 7 3750H, and the 95\% confidence intervals
overlap: [25.94, 26.06]~ms for Apple M3 and [25.54, 27.26]~ms for
Ryzen. The corresponding ARM/x86 mean-time ratio is $0.98\times$,
which should not be interpreted as a meaningful timing advantage for
either platform.

Performance stability is also workload-dependent. On Fibonacci, the
Ryzen exhibits lower relative variance than the Apple platform. On
matrix multiplication, by contrast, the Apple M3 is substantially more
stable, with a coefficient of variation of approximately 1.2\% versus
16.7\% on the Ryzen. This suggests that the regular matmul workload is
more sensitive to platform-level frequency and thermal variability on
the x86 laptop than on the Apple SoC.

\subsection{Microarchitectural Analysis}
\label{sec:uarch}

Table~\ref{tab:uarch} presents hardware performance counter results for
the portable C profiling runs. These runs use the same workload
structure on both systems but are distinct from the hand-written
assembly implementations used for the timing and energy results.

\begin{table*}[!ht]
\centering
\small
\caption{Representative microarchitectural counter rates for matched portable C \texttt{-O0} profiling runs. Apple counters were collected with Instruments CPU Counters; Ryzen counters were collected with Linux \texttt{perf stat}.}
\label{tab:uarch}
\begin{tabularx}{\textwidth}{lXXXX}
\toprule
Counter & Apple fib(40) & Ryzen fib(40) & Apple matmul & Ryzen matmul \\
\midrule
Measured window              & 37.23~s & 32.43~s & 37.87~s & 38.94~s \\
Instructions per cycle (IPC) & 3.74 & 2.49 & 5.43 & 2.83 \\
Instructions per second      & $6.61 \times 10^9$ & $9.45 \times 10^9$ & $9.41 \times 10^9$ & $1.12 \times 10^{10}$ \\
Branch misprediction rate    & 0.00067\% & 0.095\% & 0.097\% & 0.406\% \\
L1D miss/load indicator      & 0.00046\% & 0.0015\% & 4.93\% & 7.31\% \\
\bottomrule
\end{tabularx}
\end{table*}

The counter data show that the Apple M3 executes both portable-C
profiling workloads with substantially higher IPC than the Ryzen 7
3750H. For Fibonacci, the Apple run reaches 3.74 IPC versus 2.49 IPC on
Ryzen; for matrix multiplication, the corresponding values are 5.43 and
2.83. The Ryzen system nevertheless retires more instructions per
second in these profiling runs because its measured cycle rate is
higher during the runs. This helps explain why higher IPC alone does
not determine the wall-clock result.

Branch behavior is favorable on both systems, but the Apple counters
show lower misprediction rates. The difference is especially large for
Fibonacci, where the Apple run records a branch misprediction rate of
0.00067\%, compared with 0.095\% on Ryzen. Matrix multiplication has a
higher branch-misprediction rate on both systems, but the rates remain
below 0.5\%, indicating that branch prediction is not the dominant
source of cost for that workload.

The L1 data also separate the two workloads. Fibonacci has negligible
L1D load-miss rates on both systems, while matrix multiplication places
clearer pressure on the data-cache hierarchy. The Apple M3 profiling
run records a 4.93\% L1D load-miss rate for matrix multiplication,
whereas the Ryzen run records 7.31\%. Because the Apple denominator is
derived from integer load instructions and the Ryzen denominator from
reported L1 data-cache loads, these percentages should be interpreted
as comparable workload indicators rather than as perfectly identical
hardware definitions.

\subsection{Energy Efficiency}
\label{sec:energy_results}

\begin{figure*}[!ht]
\centering
\includegraphics[width=\textwidth]{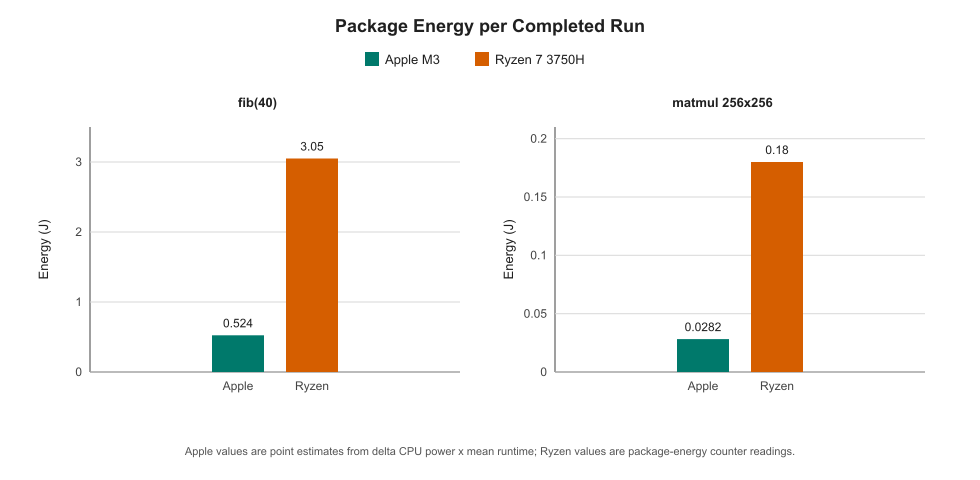}
\caption{Energy per completed benchmark run. Apple M3 values are CPU-energy point estimates derived from dynamic CPU power sampling, while Ryzen values are direct package-energy counter readings. The Ryzen/M3 energy ratios are approximately $5.82\times$ for \texttt{fib(40)} and $6.38\times$ for matrix multiplication.}
\label{fig:energy_per_run}
\end{figure*}

Table~\ref{tab:energy} and Figure~\ref{fig:energy_per_run} report the
energy per benchmark run. On Apple, these values are estimated from
dynamic CPU power sampling; on Ryzen, they are measured directly from
the \texttt{energy-pkg} package-energy counter.

\begin{table}[!ht]
\centering
\small
\caption{Energy per benchmark run (assembly implementations).}
\label{tab:energy}
\begin{tabularx}{\columnwidth}{lXXX}
\toprule
Benchmark & Apple M3 CPU estimate (J) &  Ryzen package (J) & Ryzen/M3 ratio \\
\midrule
fib(40)               & $\approx 0.5241$ &  $3.05$ &  $5.82\times$ \\
matmul $256\times256$ & $\approx 0.0282$ &  $0.18$ &  $6.38\times$ \\
\bottomrule
\end{tabularx}
\end{table}

To make the energy results easier to compare across platforms,
Table~\ref{tab:normalized_efficiency} reports normalized workload-level
metrics. For \texttt{fib(40)}, the normalization unit is one recursive
function invocation in the naive recursion tree, giving
$2F_{41}-1 = 331{,}160{,}281$ calls. For matrix multiplication, the
normalization unit is one integer arithmetic operation, with
$2N^3 = 33{,}554{,}432$ multiply/add operations for $N=256$. The table also reports the energy-delay product (EDP) for
each completed benchmark run.

\begin{table*}[!ht]
\centering
\small
\caption{Normalized workload-level efficiency metrics for the assembly benchmarks. Apple energy values are CPU-energy estimates, while Ryzen values are package-energy measurements.}
\label{tab:normalized_efficiency}
\begin{tabularx}{\textwidth}{llXXXX}
\toprule
Benchmark & Platform & Throughput & Energy/unit & Work/J & EDP (J\,s) \\
\midrule
fib(40) & Apple M3 & $5.67 \times 10^8$ calls/s & 1.58 nJ/call & $6.32 \times 10^8$ calls/J & 0.306 \\
fib(40) & Ryzen 7 3750H & $6.97 \times 10^8$ calls/s & 9.21 nJ/call & $1.09 \times 10^8$ calls/J & 1.45 \\
matmul $256\times256$ & Apple M3 & $1.29 \times 10^9$ ops/s & 0.840 nJ/op & $1.19 \times 10^9$ ops/J & $7.33 \times 10^{-4}$ \\
matmul $256\times256$ & Ryzen 7 3750H & $1.27 \times 10^9$ ops/s & 5.36 nJ/op & $1.86 \times 10^8$ ops/J & $4.75 \times 10^{-3}$ \\
\bottomrule
\end{tabularx}
\end{table*}

Although the Ryzen is faster on Fibonacci and statistically tied with
the M3 on matrix multiplication, the Apple platform uses far less
measured or estimated processor energy per completed workload in both
cases, as shown in Figure~\ref{fig:energy_per_run}. The gap is
substantial: the Ryzen/M3 energy ratio is approximately $5.82\times$
on Fibonacci and $6.38\times$ on matrix multiplication.
The normalized values in Table~\ref{tab:normalized_efficiency} show the
same pattern. The Ryzen system completes more Fibonacci calls per
second, consistent with its lower wall-clock time, but the Apple M3
performs about $5.82\times$ more recursive calls per joule. For matrix
multiplication, throughput is essentially tied, while the Apple M3
performs about $6.38\times$ more integer operations per joule and has a
substantially lower energy-delay product.

\begin{figure*}[!ht]
\centering
\includegraphics[width=0.85\textwidth]{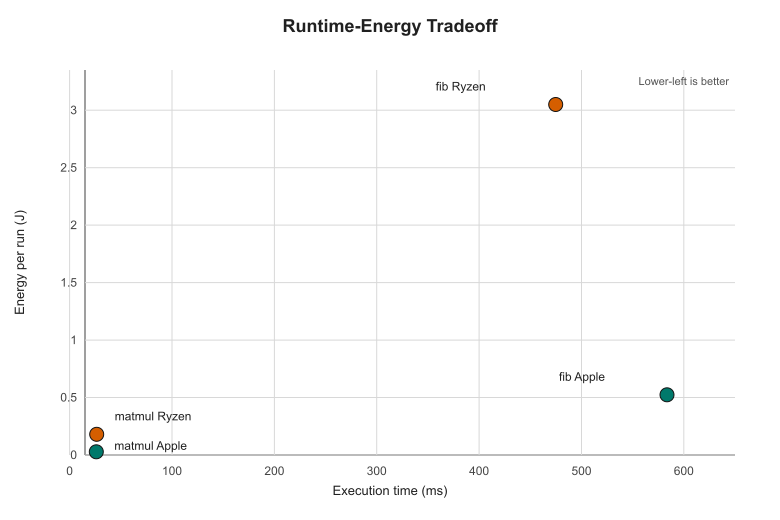}
\caption{Runtime-energy tradeoff for the two assembly benchmarks. Lower values on both axes indicate better performance and lower energy-to-solution. The Ryzen system improves Fibonacci runtime but consumes substantially more package energy, while the Apple M3 occupies the lower-energy operating point for both workloads.}
\label{fig:runtime_energy_tradeoff}
\end{figure*}

This is the central empirical result of the study. For the tested
mobile-class laptop platforms, the Apple M3 delivers markedly better
energy efficiency than the Ryzen 7 3750H, even though the difference in
raw execution time is modest and workload-dependent. The combined
runtime--energy tradeoff is summarized in
Figure~\ref{fig:runtime_energy_tradeoff}. A likely
explanation is not a single ISA-level factor, but rather the combined
effect of a newer fabrication process, lower platform power overhead,
aggressive SoC-level power management, and broader architectural
integration on the Apple system.

\paragraph{Limitations.}
The present study compares two real laptop platforms rather than two
silicon-matched processors. The systems differ in year, fabrication
process, operating system, thermal behavior, and energy-measurement
methodology. In addition, the benchmark suite is intentionally small
and focuses on two assembly workloads chosen to expose different
bottlenecks. The conclusions should therefore be read as workload- and
platform-specific rather than as exhaustive statements about either ISA
family.

\subsection{Summary of Findings}
\label{sec:summary}

The experimental results support the following conclusions:

\begin{enumerate}
  \item On raw execution time, the Ryzen 7 3750H holds a decisive
        advantage on the branch-heavy Fibonacci benchmark, with
        non-overlapping 95\% confidence intervals, while matrix
        multiplication shows no decisive time advantage for either
        platform in the present measurements.
  \item On measured or estimated processor energy per completed workload, the Apple M3 is
        substantially more efficient: the Ryzen/M3 energy ratio is
        approximately $5.82\times$ on Fibonacci and $6.38\times$ on
        matrix multiplication. The normalized calls-per-joule and
        operations-per-joule metrics show the same advantage.
  \item Runtime variability is workload-dependent: the Ryzen is more
        stable on Fibonacci, whereas the Apple platform is much more
        stable on matrix multiplication.
  \item The matched portable-C counter measurements show that the Apple
        M3 sustains higher IPC on both profiling workloads, while the
        Ryzen system retires more instructions per second because of
        its higher measured cycle rate.
  \item The counter data also confirm that the two workloads stress
        different parts of the processor: Fibonacci remains primarily
        control-flow and call intensive, whereas matrix multiplication
        places greater pressure on the data-cache hierarchy.
\end{enumerate}

Overall, these results show that the tested ARM- and x86-64-based
laptop platforms occupy different operating points in the
performance--energy tradeoff space. The Ryzen platform retains an
advantage on the branch-heavy recursive workload, but the Apple M3
achieves substantially lower energy per completed task across both
benchmarks. In that sense, the results support the broader view that
ARM-based laptop-class systems are no longer limited to low-power
niches, but are competitive in general-purpose computing when
energy-to-solution is a primary metric.

\section{Conclusion}

This study combined architectural analysis with native benchmarking to compare an
Apple M3 platform and an AMD Ryzen 7 3750H platform on two hand-written assembly
workloads. The results show a clear tradeoff rather than a universal winner. The
Ryzen platform holds a timing advantage on the branch-heavy Fibonacci benchmark,
whereas matrix multiplication shows no decisive runtime advantage for either system
in the present measurements. In contrast, the Apple platform delivers a much lower
energy-to-solution on both workloads.

These findings are consistent with the view that modern ARM-based laptop systems are
competitive in general-purpose computing when energy efficiency is treated as a
primary metric. At the same time, the results should not be interpreted as a pure
ISA-level verdict. The measured differences reflect not only instruction-set design,
but also process technology, SoC integration, cache organization and operating-system
behavior.

The architectural discussion and cross-platform counter data help explain the
observed behavior. The matched portable-C profiling runs show higher IPC on the
Apple M3, but also show that the Ryzen platform retires more instructions per
second because of its higher measured cycle rate. Fibonacci places greater
emphasis on control flow and function-call handling, where the Ryzen platform
performs strongly in the assembly timing measurements, while matrix
multiplication places greater pressure on the memory hierarchy. The Apple
system, meanwhile, appears to benefit from a platform design that sustains far
lower energy per completed task across both workloads.

Future work may extend the benchmark suite with additional workload classes,
incorporate more closely matched platforms, and broaden the PMU analysis to
optimized C, hand-written assembly, and additional counter groups in order to
refine the distinction between ISA-level and platform-level effects.

\printbibliography

\end{document}